\documentstyle[11pt,newpasp,twoside,epsf]{article}
\def\edcomment#1{\iffalse\marginpar{\raggedright\sl#1\/}\else\relax\fi}
\reversemarginpar

\begin{document}
\title{Introductory Review to the Historical Development of Modern Cosmology}
\author{Vicent J. Mart\'{\i}nez}
\affil{Observatori Astron{\`o}mic, Universitat de Val{\`e}ncia,
E46100-Burjassot, Val{\`e}ncia, Spain}

\begin{abstract}
This talk introduces the contributions of the lecturers 
to the international summer school 
``Historical Development of Modern Cosmology''
held at the Universidad Internacional Men\'endez Pelayo, Valencia, Spain,
during September 18-22 2000. 
\end{abstract}

\section{Prelude}

This school on the Historical Development of Modern Cosmology is
taking place when the University of Valencia celebrated 500 years
of uninterrupted academic and scientific activity. Let me start
the introductory talk with a prelude regarding some astronomical
events that happened at the very beginning of the University's
existence. A few years after the foundation of the University, Jeroni
Muny\'os, Professor of Astronomy, Mathematics and Hebrew at this
University, studied the famous Type I Supernova within the
constellation of Cassiopeia, which appeared in 1572 (Tycho's Supernova)
and wrote a small book commissioned by the king Felipe II. The
book (see Fig. 1) was entitled
\begin{quote}
{\it Libro del nuevo Cometa, y del lugar donde
se hazen; y como se ver{\'a} por las Parallaxes, qu{\'a}n
lexos est{\'a}n de tierra...\\
(Book on the new Comet, and the place where it can be found, and as
it can be seen from its
parallax how far away it is from Earth...)
}
\end{quote}
Muny\'os knew that the new star could be used as an observational
evidence against the Aristotelian concept of the immutability of
the Cosmos, and of course, this work brought him some problems
with the political and ecclesiastic authorities of the time.
Nevertheless, Tycho Brahe knew Muny\'os's work well and, in a book
(see Fig. 1) in which Tycho discusses the work by other astronomers on
the new star, he includes a chapter devoted to Muny\'os's book 
(Navarro Brot\'ons and Rodr\'{\i}guez Galdeano 1998).

\begin{figure}[t]
\begin{center}
\vspace{2.cm}
\epsfxsize=4.95cm
\epsfxsize=7.26cm
\caption{Left: The cover of the book on the `New Comet' published
in 1573 by Jeroni Muny\'os commissioned by the Spanish king Felipe
II. Valentia, Pedro de Huete, 1573. Courtesy of the Biblioteca Valenciana,
Biblioteca Nicolau Primitiu, Val\`encia.
Right: Cover of the book {\it Astronomiae Instauratae Progymnasmata}, 1648,
(Impensis Ioannis Godophredi Sch\"onwetteri) 
in which Tycho Brahe discussed work 
on the new star (Tycho's Supernova in Cassiopeia) by many
other European astronomers. Courtesy of the Biblioteca Hist\`orica,
Universitat de Val\`encia.}
\end{center}
\end{figure}

\section{
What was the aim of the school?
}

In the book by H. Kragh (1996), we can read:

\begin{quote}{
``In spite of cosmology's amazing development in this century and
the strong scientific and public interest in the new science of
the universe, only very little is known of how this development
took place."}
\end{quote}

Also in  his book, Kragh quotes Y. Zel'dovich and I. Novikov (1983) from
their book {\it Structure and Evolution of the Universe}:
\begin{quote}{
``The history of the universe is infinitely more interesting than
the history of the study of the universe."
}
\end{quote}

Undoubtedly, this view is still shared by many professional
astronomers. Nevertheless, it is our responsibility to provide 
future generations of scholars in this field with a truthful
account of how the present cosmological paradigm has been
constructed from the contributions of many people. This school
points in this direction and these proceedings will probably
help to increase the general knowledge of how our
concept of the universe has been changing during the last 80
years, a period in which cosmology has experienced a dramatic
development. Sometimes controversy between alternative theories
was present in this history. The Steady State model
and the Big Bang model, with once-strong arguments supported by prestigious 
defenders on both sides, is
a good example of controversial science. These arguments are
nicely explained in the chapters
by Kragh (p. 157), Gold (p. 171), and Narlikar (p. 175).

The new cosmological ideas soon permeated society. It is well
known that the term ``Big Bang" was coined ironically by one of the
founders of the Steady State theory, Sir Fred Hoyle, during a BBC
radio interview. Two talks of this school will be devoted to the
popularization of Cosmology: Ten (p. 309) focuses on the 
ideas at the beginning of the twentieth century and Hawkins (p.
325) discusses the modern methods of disseminating 
discoveries about the universe.

\section{The beginning of the story}

How did this history begin? We could start from many different
significant moments and discoveries. I like to remember the comet
hunters during the eighteenth and the nineteenth century. 
The catalog of extended objects, compiled by Charles Messier (1730-1817)
to avoid their being confused with new comets, is the first listing of
nebulae widely referenced by later astronomers. 
Lord Rosse discovered
the spiral structure of some of these nebulae using his 1.8-meter
reflecting telescope at Birr Castle (Ireland). His drawing of M51
(see Fig. 2)
has a remarkable resemblance with modern images of this spiral
galaxy.

\begin{figure}
\begin{center}
\begin{minipage}{0.85\textwidth}
\vspace{2.cm}
\end{minipage}
\end{center}
\caption{The drawing of the Whirlpool galaxy (M51) by William
Parsons (Lord Rosse) performed in 1845 while observing with the
Leviathan of Parsonstown. Courtesy of the Birr Scientific and
Heritage Foundation, Birr Castle, Birr, Ireland.}
\end{figure}

The nature of the nebulae and their possible extragalactic character was
one of the items discussed in the debate on the distance scale of the
universe, held in Washington on 26 April 1920 between Harlow Shapley
of Mt. Wilson (soon after director of Harvard College Observatory) and
Heber D. Curtis of Lick (later director of Allegheny Observatory).
Shapley placed the sun far from the center of a galaxy 300\,000 light
years in diameter and the spiral nebulae inside it.  Curtis instead
put the sun at the center of a 30\,000 light year galaxy and the nebulae
outside it as concentrations of stars similar to the Milky Way.  It
is clear that Shapley was entirely right about the location of the sun,
and more nearly right about the distance scales, but Curtis was right
about the nebulae.

The definitive clarification of the controversy came with Edwin
Hubble in 1925 (but see also the contributions by Ernst {\"O}pik
and Knut Lundmark explained by Einasto on p. 85 and by Trimble on p. 
375). Hubble discovered 
Cepheid variable stars in NGC 6822 and in Andromeda (1925a,b). 
He estimated the 
distance to Andromeda 
using the period--luminosity relationship introduced by
Henrietta Leavitt in 1908. Although still underestimated, the
distance obtained by Hubble, 900\,000 light years, was the end of
Shapley's universe. This measurement, together with the
morphological classification of galaxies and the discovery of the
universal cosmic expansion, were the major contributions of Hubble
to the field. Moreover, Hubble's law ---the redshift--velocity
relation--- is one of the most important breakthroughs in science
that have happened during the twentieth century (see the contribution by
Christianson on p. 145.)

\begin{figure}[h]
\begin{center}
\epsfxsize=6.05cm
\begin{minipage}{\epsfxsize}\epsffile{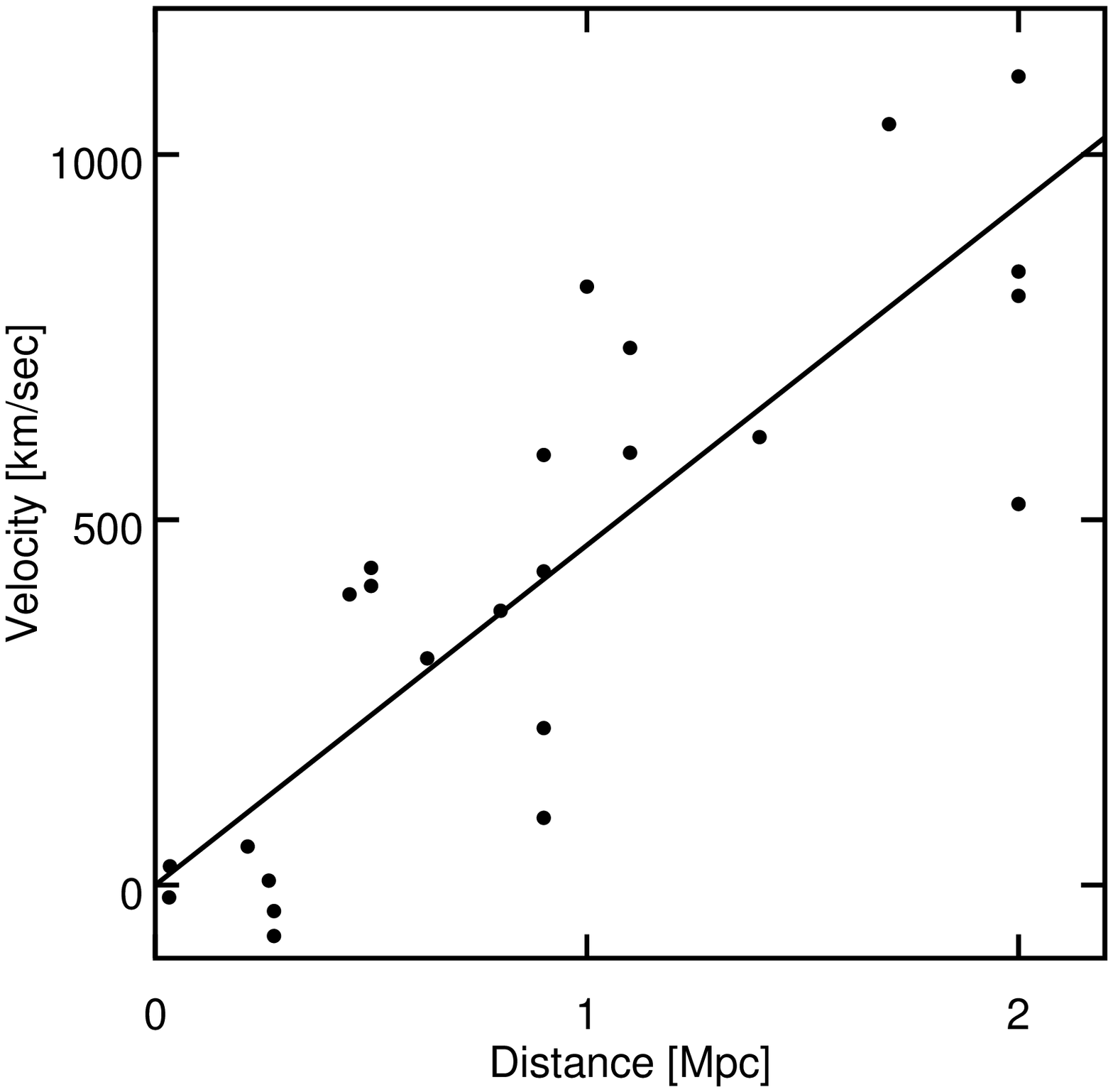}\end{minipage}
\epsfxsize=6.534cm
\begin{minipage}{\epsfxsize}\epsffile{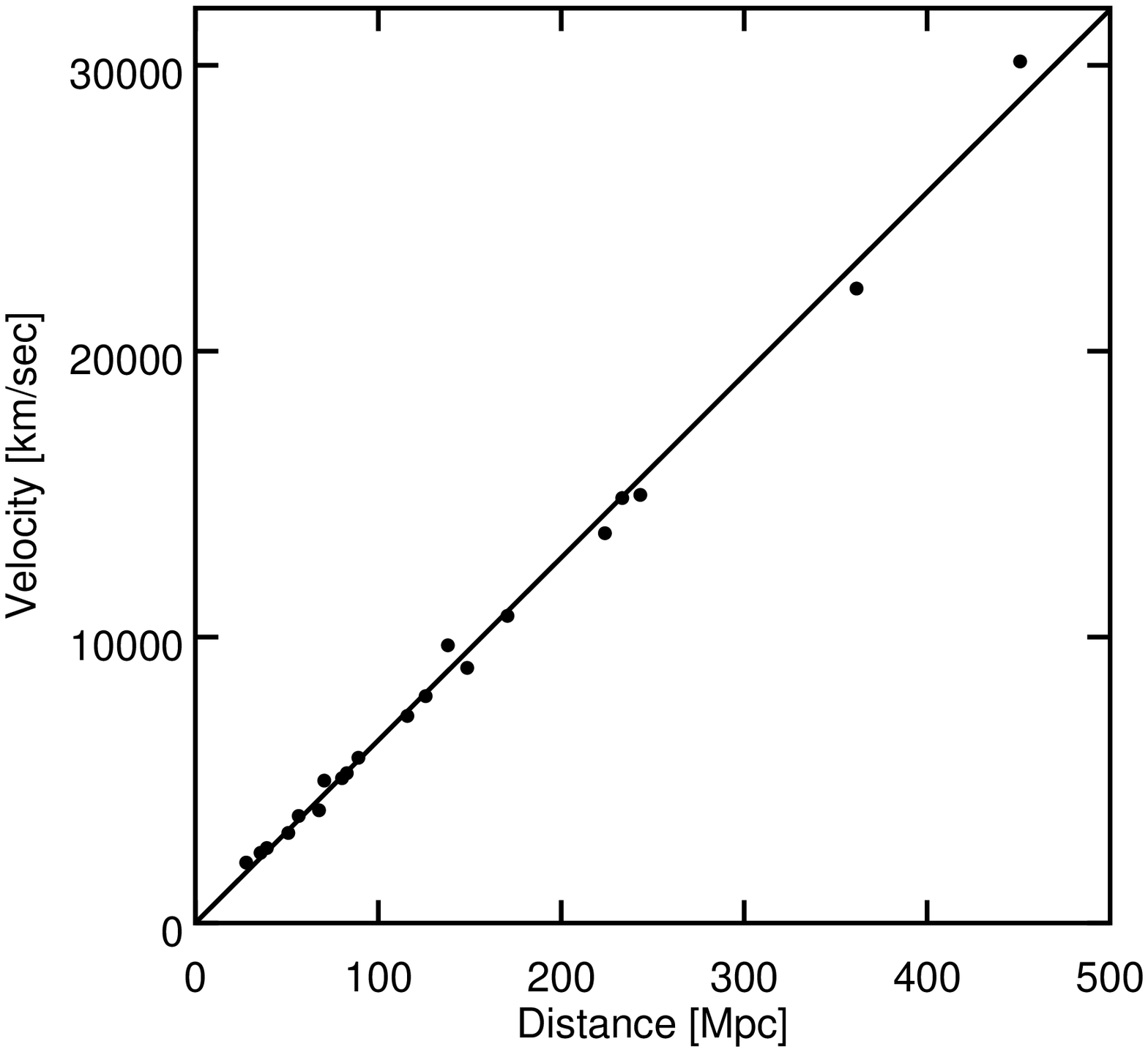}\end{minipage}
\caption{The velocity--distance relation. In the left 
panel are plotted the data shown by Hubble in his 1929 paper
together with his linear fit (hard to believe by modern standards).
In the right panel are data for Type Ia SNe obtained
by Riess, Press, and Kirshner (1996).
\copyright Edward L. Wright (UCLA), used with permission. }
\end{center}
\end{figure}

During the eighteenth century Immanuel Kant and Johann Lambert
conceived the idea of a hierarchical universe of stars clustered
into large systems (galaxies), which in turn are clustered into
larger systems, and so on. Hierarchical clustering as a solution
of Olber's paradox was proposed by John Herschel and Richard
Proctor in the nineteenth century. The same idea was taken up in
the twentieth century by Carl Charlier. The first talk after this
introduction deals precisely with Olber's paradox (Hoskin p. 11).

\section{Modern cosmology is born}

Despite the beauty of the hierarchical universe, this idea
was not widely accepted. In fact, 
one of the tenets of modern cosmology was the
Cosmological Principle. The principle was first formulated by E. Milne in 1933
as the assumption that the large-scale unverse is spatially
homogeneous and isotropic. Many authors in this volume regard the
publication of the theory of General Relativity by Albert
Einstein in 1915 (Einstein 1915a,b) 
as the birth of modern cosmology. As soon as 
1919, the first experimental test of the theory of General
Relativity was performed in the expeditions led by Sir Arthur Eddington
and Andrew Crommelin to
measure, during a total solar eclipse, the bending of light rays
from background stars by the sun (see Coles on p. 21)

In 1917 Einstein applied his equations of General Relativity
to the description of the universe, assuming the Cosmological
Principle. This assumption was adopted mainly for simplicity, and
needed to be confirmed by observations. When the first maps of the
distribution of galaxies, both in projection and later in redshift
space, became available, a clumpy distribution was revealed. In the 1970s,
Peebles and co-workers used
statistical descriptors such as the two-point correlation function
to conclude that the small-scale clustering
of galaxies was compatible with a fractal distribution, and
therefore a transition to homogeneity had to happen 
if the Cosmological Principle was to be preserved.  Observations of
redshifts of galaxies out to 15\,000 km/sec and beyond, the isotropy of
the X-ray background, and especially of the microwave background
eventually confirmed the principle.  But data sets from the 1960s to
the 1980s showed structure nearly as large as the survey volumes and
left the issue open in the minds of some astronomers.  
The arrangement of matter in the universe, how galactic structures
have been formed and how they can be statistically described 
are the main topics of the contributions by Peebles (p.
201), Jones (245), and Icke (p. 337). In addition, numerical 
simulations have played an important role in recent
years in exploring the consequences of different scenarios for the
formation and evolution of large scale structure in the 
universe (see Yepes on p. 355).

\subsection{The cosmological constant}

Einstein introduced the cosmological constant, $\Lambda$, in his
field equations to guarantee a static universe. Some years later,
and after the discovery by Hubble of the expansion of the
universe, he considered this introduction as his biggest blunder (see Heller
on p. 121). Nevertheless, recent observations 
(Riess et al. 1998, Perlmutter et al. 1999) of very distant Type
Ia Supernovae have resurrected the cosmological constant, interpreting 
it as the imprint of an exotic kind of energy that could be a
possible explanation for a universe with accelerated expansion (Silk p. 109).
The solutions of the Einstein equations obtained by the Russian
mathematician Alexander Friedmann (1922) and the Belgian mathematician 
(and priest)
George Lema\^{\i}tre (1927) (see Blanchard on p. 237) were the
basis for the standard cosmological model of a dense, hot, and
expanding universe. Some of the more influential contributions in
the 1930s were the Einstein-de Sitter flat model (1931), and the
papers by Robertson (1933) and Tolman (1934). The problem of the
singularity at the beginning of the history of the universe is
discussed at length by Heller (p. 121).

\subsection{Primordiual nucleosyntesis}

During the first few minutes after the Big Bang the formation of
light elements took place. One of the
first works on this ``primordial nucleosynthesis" is the famous
$\alpha\beta\gamma$ paper published in 1948 (Alpher, Bethe, and Gamow
1948). During the fifties and sixties, different
calculations of the relative proportions of hydrogen, helium,
deuterium and lithium within the framework of the Big Bang model
showed an excellent agreement with the present-day observed abundances. This
was the first theoretical prediction of the new cosmological model
confirmed by the observations.

\subsection{Cosmic microwave background radiation}

In 1965, the American radio astronomers Arno Penzias and Robert
Wilson detected the cosmic microwave background (CMB) radiation.
The existence of this diffuse radiation is naturally explained in
the Big Bang model. When the age of the universe was about
300\,000 years, the universe had cooled enough to make possible
the formation of stable hydrogen atoms for the first time (epoch
of recombination). 
At this stage, the universe becomes transparent, because the 
electromagnetic radiation and matter decouple. Indeed Alpher and Herman (1950)
had predicted that such radiation should exist at a temperature of 
about 5K.

The reported detection by
Penzias and Wilson was published in The Astrophysical Journal.
In a paper published in the same issue of the journal,
Robert Dicke, James Peebles, Peter Roll, and
David Wilkinson explained the cosmological origin of this
radiation. Without any doubt, this discovery was the strongest
observational evidence supporting the Big Bang model. The news
appeared on the first page of The New York Times on 21 May
1965. Since then, the cosmic microwave radiation, as an
observed fossil of the early universe, has been one 
of the main targets of observational cosmology, carried out wiht
the COBE (COsmic Background Explorer) satellite, the BOOMERANG balloon
experiment, and from the ground. After
1965, the Big-Bang model becomes the generally accepted
cosmological model, overshadowing the Steady State theory. The
contributions by Novikov (p. 43) and by Silk (p. 109) deal with the
CMB observations, their discovery, their analysis and 
their cosmological implications,
for example the flatness of the universe.

\section{The model}

Summarizing, the hot Big Bang model accounts for three
observational facts:
\begin{itemize}
\item
The current expansion of the universe.
\item
The cosmic microwave background radiation.
\item
The relative abundance of light chemical elements.
\end{itemize}


These successes are important, but the model needs several more
ingredients to provide a complete picture of the origin and evolution
of the universe:
\begin{enumerate}
\item {\bf Cosmological parameters.}

Alan Sandage (1970) once described cosmology as the search
for two numbers, the Hubble constant and the deceleration parameter 
(the latter is, for a model with zero
cosmological constant, half the density parameter $\Omega$.) A few
more numbers are considered today as fundamental parameters in
modern versions of cosmological models (Silk p. 109; Trimble p. 375;
see also Rees 2001).

\item {\bf Inflation}. The successful description of the universe, provided
by the Big Bang model, leaves many questions open such as
{\it why is the universe so homogeneous, so isotropic, and so flat?}.
A 1990s answer is the idea that the very early universe
passed through a period of exponential expansion, called inflation
(Kolb p. 295; Bonometto p. 219).
\begin{figure}[h]
\begin{center}
\begin{minipage}{0.7\textwidth}
\plotone{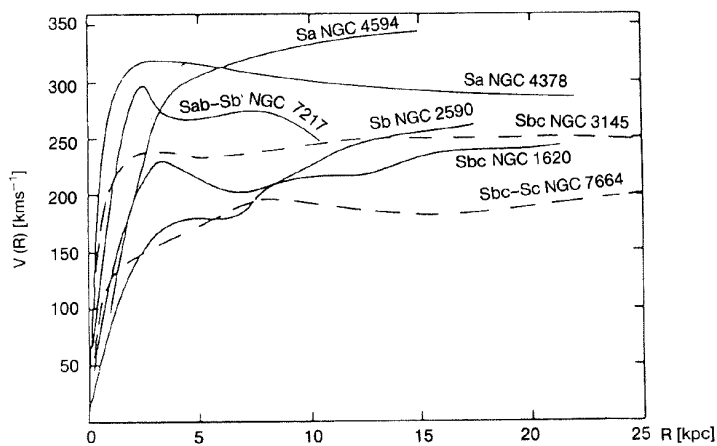}
\end{minipage}
\end{center}
\caption{The galactic rotation curves for seven spiral galaxies.
The flatness of these curves at large distances from the galactic
center is explained by the presence of huge dark halos surrounding
the galaxies. Reproduced, with permission, from Rubin, Ford, and
Thonnard 1978, ApJ, 225, L107.}
\end{figure}

\item {\bf Dark matter.}
One of the first mentions of missing mass or dark
matter came from Zwicky (1933), who had measured the velocity dispersion
in the Coma cluster and applied the virial theorem. The acceptance
of the dark matter was a very slow process (see the
contribution by van den Bergh on p. 75.) In any case, it took
longer than the more recent general acceptance of the existence of
dark energy (see Kolb on p. 295 and Bonometto on p. 219.) The
contribution of the Tartu astronomers to the dark matter problem
was remarkable (Einasto p. 85.) The evidence provided by the
rotation curves in spiral galaxies was overwhelming (Rubin et al.
1978). These curves
indicated that spiral galaxies contain at least 10 times more dark
matter than luminous matter (see Fig. 4). Different kinds of dark
matter enter as contributors for the total density of the universe
(Trimble 1987). 
The role of neutrinos is explained in this volume by Valle (p 275.)

\end{enumerate}

\section{Looking forward}

Cosmology is a science driven by observations and
observations are driven today by technological developments
(Longair p. 55.) The technologies to be developed in the 21st
century will provide deeper insights into the universe. Many
problems are still open. Some of them are part of what Bonometto (p. 219) 
calls post-modern cosmology, which addresses questions such as
{\it what happened before the big bang?} or {\it are we living in
one universe amongst many others?} Moreover, 
the present model seems too elaborate, requiring fine-tuned values for the
parameters and many different contributions to the total density
(Kolb p. 295.) In any case the nature of dark matter and dark
energy is still unknown. In my opinion, although we can be proud
of the knowledge of the universe acquired during the last
years, honestly (and modestly) we have to admit that it is only
the tip of the iceberg.

\acknowledgements
{I would like to thank Bernard Jones, V\'{\i}ctor Navarro Brot\'ons, 
Mar\'{\i}a Jes\'us Pons-Border\'{\i}a,
and Virginia Trimble for valuable comments and suggestions.
This work was supported by the Spanish MCyT project AYA2000-2045.
}


\begin{references}
\reference Alpher, R. A., Bethe, H., and Gamow, G. 1948, Phys.Rev. 73, 803
\reference Alpher, R. A., and Herman, R. C. 1950, Rev.Mod.Phys. 22, 153
\reference Dicke, R. H., Peebles, P.J.E., Roll, P. G., and
Wilkinson, D. T. 1965, \apj \, 142, 414
\reference Einstein, A. 1915a Preuss. Akad. Wiss. Berlin,
Sitzber., 778
\reference Einstein, A. 1915b Preuss. Akad. Wiss. Berlin,
Sitzber., 844
\reference Einstein, A. 1917 Preuss. Akad. Wiss. Berlin,
Sitzber., 142
\reference Einstein, A., and de Sitter, W. 1931, Proc.Nac.Acad.Sci 18, 213
\reference Friedmann, A. A. 1922, Zeitschr. f{\"u}r Phys. 10, 377
\reference Hubble, E. P. 1925a, \apj \, 62, 409
\reference Hubble, E. P. 1925b, Observatory 48, 139
\reference Hubble, E. P. 1929, Proc.Nac.Acad.Sci 15, 168
\reference Kragh,  H. 1996, Cosmology and Controversy. The Historical
Development of Two Theories of the Universe
(Princeton: Princeton University Press)
\reference Lema\^{\i}tre, G. 1927, Ann. Soc. Scient. Brux., 47A, 49
\reference Milne, E. A. 1933, Zeitschr. f{\"u}r Astrophys. 6, 1
\reference Navarro Brot\'ons, V. and Rodr\'{\i}guez Galdeano, E. 1998,
Matem\'aticas, cosmolog\'{\i}a y humanismo en la Espa\~na del siglo XVI.
Los ``Comentarios al segundo libro de la Historia Natural de Plinio" de
Jer\'onimo Mu\~noz (Valencia: Ins\-ti\-tuto de Estudios Documentales e Hist\'oricos
sobre la Ciencia). 
\reference
Penzias, A. A. and Wilson, R. W. 1965,  \apj \, 142, 419
\reference Perlmutter, S. et al. 1999, \apj \, 517, 565
\reference Rees, M. J. 2001, Just Six Numbers: 
The Deep Forces that Shape the Universe (New York: Basic Books)
\reference 
Riess, A. G., Press, W. H., and Kirshner, R. P. 1996, \apj \, 473, 88
\reference Riess, A. G. et al. 1998, \aj \, 116, 1009
\reference
Robertson, H. P. 1933, Rev.Mod.Phys. 5, 62.
\reference
Rubin, V. C., Ford, W. K., and Thonnard, N. 1978, ApJ 225, L107
\reference
Sandage, A. 1970, Physics Today 23, 31
\reference
Tolman, R.C. 1934, Relativity, Thermodynamics, and Cosmology (Oxford: Oxford
University Press)
\reference Trimble, V. 1987, \araa \, 25, 425 
\reference Zel'ldovich, Ya. B., and Novikov, I. D. 1983, Relativistic 
Astrophysics. II: The Structure and Evolution of the Universe (Chicago:
Chicago University Press)
\reference Zwicky, F. 1933, Helv. Phys. Acta 6, 110
\end{references}
\end{document}